\documentclass[final,3p,times,twocolumn]{elsarticle}
 \biboptions{comma,sort&compress}
\usepackage{here}
 \usepackage{graphicx}
  \usepackage{epsfig}

\usepackage{amsthm}
\usepackage{amsmath}
\usepackage{amssymb}
\usepackage{amsfonts}
\usepackage{enumerate}
\usepackage{rotating}

\theoremstyle{definition}

\theoremstyle{remark}

\theoremstyle{plain}

\newcommand{\calo}{\mathcal{O}}

\def\nin{\noindent}
\def\beq{\begin{equation}}
\def\eeq{\end{equation}}
\def\bea{\begin{eqnarray}}
\def\eea{\end{eqnarray}}

\journal{Nuc. Phys. (Proc. Suppl.)}

\begin{document}

\begin{frontmatter}



\title{Quark scalar current correlator in AdS/QCD approach}

 \author{Herintsitohaina M. Ratsimbarison}
  \address{Institute of High Energy Physics - Madagascar,\\
  Q 208, Faculty of Science Building,\\
  Antananarivo University\\
  (in collaboration with Prof. S. Narison)}
\ead{hery@refer.mg}



\begin{abstract}
\noindent
We use a holographic model to compute the quark scalar current correlator. Already used in other contexts, our model consists in a bulk field theory defined on a deformed Anti de Sitter space and allows to derive analogous results to those of the soft wall model. One issue of such computations is to understand relevant holographic physical constraints which give predictive QCD results rather than to make the inverse step as done until now.
\end{abstract}




\end{frontmatter}

\section{Introduction}
Initiated by the AdS/CFT correspondence, the holographic approach is on the way to provide theoretical tools to do nonperturbative QCD. The holographic principle relates the generating functional of some field theory of interest with the partition function of a dual theory defined on a bulk space with boundary conformally identical to the spacetime of the field theory \cite{edwi98}:
\begin{align}
 \left\langle0| \mathcal{T}exp(i\int\phi_0\calo)|0\right\rangle_{boundary} &=& \int_{\phi|_{M_d} = \phi_0}D\phi\; e^{iS_{bulk}[\phi]}, \label{holcor}
\end{align}
where the boundary restriction $\phi_0$ of the bulk field $\phi$ is the source of the operator $\calo$ defined on the boundary theory.\\
Soft wall and hard wall models are among the simplest models which allow to describe QCD in this manner. Following previous works on the gluonium case \cite{hera09}, the present paper deals with the quark scalar current correlator and compares the soft wall model with those introduced in \cite{oavz07}.

\section{The quark scalar current correlator}

For light quarks with Dirac fields $\psi$, the quark scalar current $\bar{\psi}\psi$ and its associated 2-point correlation function 
\begin{eqnarray*}
\left\langle \Omega \left|\mathcal{T} (\bar{\psi}\psi)(x)(\bar{\psi}\psi)^{\dagger}(0)\right| \Omega\right\rangle
\end{eqnarray*}
allow to study masses and decay constants of light scalar mesons. \\
In QCD sum rules, its asymptotic expansion in Q$^2$ looks like
\begin{eqnarray*}
&& i\int d^4x\,e^{iqx}\,\left\langle \Omega \left|\mathcal{T} (\bar{\psi}\psi)(x)(\bar{\psi}\psi)^{\dagger}(0)\right| \Omega\right\rangle \\	
&=& \frac{3}{8\pi^2}\left[Q^2ln(\frac{Q^2}{\nu^2}) + C_pQ^2 + C_1ln(\frac{Q^2}{4\nu^2})\right]\\
&& + \frac{3}{8\pi^2}\left[\frac{C_2}{Q^2} + \frac{C_4}{Q^4} + \frac{C_6}{Q^6} + O(\frac{1}{Q^8})\right],	
\end{eqnarray*}
where C$_i$, i=1,2,..., are Wilson coefficients related to nonpertubative contributions from QCD condensates. One main issue of this computation is to prove the existence or not of the C$_1$ term. In principle, one expects that there is no condensate related to this term due to gauge invariance. In \cite{chetetal98}, such 1/Q$^2$ term is obtained by introducing a tachyonic gluon mass in order to improve phenomenological results on the QCD coupling constant and light quark masses. We will see that holographic models, such as the soft wall model, tend to confirm the existence of such expansion term.

\section{AdS/QCD computations}
Let us construct a theory on a five dimensional space which gives an estimate of the 2-point correlator associated to the quark scalar current. We take a modified five dimensional Anti de Sitter space with metric given in Poincar\'e coordinates by: 
\begin{eqnarray}
 && ds^2 = h(z)\frac{R^2}{z^2}\left(\eta_{\mu\nu}dx^{\mu}dx^{\nu} + dz^2\right), \\
 && \quad \quad =: g_{MN}dx^Mdx^N ,\\
 &\text{with}& ds^2 \sim ds^2_{AdS_5} = ds^2_{h(z) = 1} \text{ for z }\rightarrow 0, \label{metcond}
\end{eqnarray}
where $\eta_{\mu\nu}$ is the Minkowski metric with signature (-,+++) and R is the AdS curvature radius. The condition near z = 0 is such that our bulk space possesses conformal boundary where lives QCD-like theory described by QCD operators.
According to the ansazt (\ref{holcor}), the quark scalar current will couple to a scalar source due to Lorentz invariance of the current-source coupling term, and that the bulk field, having the source as its boundary restriction, has mass m$_5$ given by:
\begin{eqnarray*}
 m^2_5R^2 = (\Delta - p)(\Delta + p - 4) = -3.
\end{eqnarray*}
Considering a bulk theory without dilaton background like in holographic models of O. Andreev and V.I. Zakharov \cite{oavz07,snvz09}, which we will call the \emph{h-model}, we obtain the following bulk action in the present case (neglecting interaction)
	\begin{eqnarray*} 
	S[\phi] = \frac{-1}{2k}\int d^5x\sqrt{-det(g)}\;\left[g^{MN}\partial_{M}\phi\partial_{N}\phi + m_5^2\phi^2\right], 
\end{eqnarray*}
where k is the 5D Newton's constant.\\
Applying the minimal action principle for our bulk theory leads to the following equation of motion:
\begin{eqnarray}
	&&\partial_M(\sqrt{g}g^{MN}\partial_N\phi) - \sqrt{g}m_5^2\phi = 0,\\
	&\Leftrightarrow & \left[\partial_z^2 + 3\left(\frac{\partial_zh}{2h} - \frac{1}{z}\right)\partial_z - q^2 - \frac{m^2_5R^2h}{z^2}\right]\tilde{\phi} = 0; \label{hmbeom}
\end{eqnarray}
where $\tilde{\phi}$ is the Fourier transform
\begin{eqnarray*}
 \tilde{\phi}_{crit}(q,z) = \int d^4x'\;e^{-iqx'}\tilde{K}(q,z)\phi_0(x'). 
\end{eqnarray*}
We can make comparison with the soft wall model \cite{pcol087,fjug08} which uses pure AdS space and the action is given by:
\begin{eqnarray*}
 && S[\phi] = \frac{-1}{2k}\int \sqrt{g}\;e^{-(cz)^2}\left[\partial^{M}\phi\partial_{M}\phi + m_5^2\phi^2\right], \\
 && \text{c is an inverse length parameter,}
\end{eqnarray*}
with the associated bulk equation of motion
\begin{eqnarray}
 \left[\partial^2_z - 3\left(\frac{2}{3}c^2z + \frac{1}{z}\right)\partial_z - q^2 - \frac{m^2_5R^2}{z^2}\right]\tilde{\phi} = 0, \label{swqbqeom} 
 \end{eqnarray}
equivalent to the Kummer equation
\begin{eqnarray}
 \left[x\partial_x^2 + (2 - x)\partial_x - \frac{q^2 + 6c^2}{4c^2}\right]Ku(x) = 0,
\label{eqkum}
\end{eqnarray}
under the change of variable x = c$^2$z$^2$ and Ku(x) = z$^{-3}\tilde{\phi}$(z).\\ 
Now, the 2-point function of the quark scalar current is derived from the second functional derivative of the critical bulk action, and we have:
\begin{eqnarray*}
&& \left\langle \Omega |\mathcal{T} (\bar{\psi}\psi)(x_1)(\bar{\psi}\psi)^{\dagger}(x_2)| \Omega\right\rangle \\
&=& \frac{(-i)^2i\delta^2S[\phi_{crit}]}{\delta\phi_0(x_1)\delta\phi_0(x_2)},\\
\Pi_{q\bar{q}} &=&	\int d^4x\;e^{iqx}\left\langle \Omega |\mathcal{T} (\bar{\psi}\psi)(x)(\bar{\psi}\psi)^{\dagger}(0)| \Omega\right\rangle \\
&=& \left[(\frac{1}{2k})\frac{2R^3h^{3/2}}{z^3}\tilde{K}\partial_z\tilde{K}(q,z)\right]^{z\rightarrow +\infty}_{z\rightarrow 0}
\end{eqnarray*}
where the critical field $\tilde{\phi}_{crit}$, solution of the motion equation (\ref{hmbeom}), is expressed in terms of its boundary restriction $\phi_0$(x'$\in$Boundary space) and the bulk-to-boundary propagator $\tilde{K}$(q,z) defined by:
\begin{eqnarray*}
 && \tilde{\phi}_{crit}(q,z) = \int d^4x'\;e^{-iqx'}\tilde{K}(q,z)\phi_0(x'), \\
 &&\lim\limits_{z\to 0}\tilde{K}(q,z) = 1
\end{eqnarray*}
By construction, $\tilde{K}$ is solution of the motion equation and its boundary condition is necessary in order to match $\phi$ with $\phi_0$ on the boundary space. Its use allows to compute explicitly the functional derivative with rapport to the source. However, in the soft wall model, the solution $\tilde{K}$ does not satisfy this boundary condition. Indeed, the general solution of (\ref{swqbqeom}) is given by \cite{adma48}:
\begin{eqnarray*}
  \tilde{K}(q,z) &=& z^3Ku(c^2z^2) \\
   &=& Az^3\,_1F_1(a,3,c^2z^2) \\
   &&+ Bz^3ln(c^2z^2)\,_1F_1(a,3,c^2z^2)\\
   && + B\displaystyle\sum^{+\infty}_{n=1}\frac{\Gamma(n+a)\Gamma(2)B_n(c^2z^2)^nz^3}{\Gamma(a)\Gamma(n+2)n!}\\
   && + B\frac{\Gamma(3)\Gamma(a - 1)z}{\Gamma(a)c^2},
\end{eqnarray*}
where A and B are integration constants, a = $\frac{q^2}{4c^2} + \frac{3}{2}$, and
\begin{eqnarray*}
    B_n &=& \left(\frac{1}{a} + \frac{1}{a+1} + ... + \frac{1}{a+n-1}\right) \\
   && - \left(\frac{1}{b} + \frac{1}{b+1} + ... + \frac{1}{b+n-1}\right)\\
   && - \left(1 + \frac{1}{2} + ... + \frac{1}{n}\right),
\end{eqnarray*}
so 
\begin{eqnarray*}
  \tilde{K}(q,z) &=& \frac{B}{(a-1){c}^{2}}z \\
  &&+ (A+B\ln(c^2 z^2))z^{3} + 0(z^5),
\end{eqnarray*}
and $\lim\limits_{z\to 0}\tilde{K}$(q,z) $\neq$ 1.\\
This drawback is get around by taking $\frac{\phi}{z}$ instead of $\phi$ to be dual to $\bar{\psi}\psi$ \cite{jerletal05} and for B = (a-1)${c}^{2}$. Moreover, the integration constant A can be fixed by taking normalizable solutions, and we have A = B($\Psi$(a) + 2$\gamma$ - 1).
In this case, the soft wall model gives the following estimate for the correlator at large Q$^2$ = q$^2$
\begin{eqnarray*}
 \Pi_{q\bar{q}}/C_0 &=& \frac {1}{{z}^{2}}+4{c}^{2}\left(a-1 \right)\ln\left({c}^{2}{z}^{2}\right)- c^2 \\
 && + 4{c}^{2}\left(a-1 \right)\left(\Psi(a)+  2\,\gamma - \frac{1}{2}\right) + O(z^2)\\
 &=& Q^2ln(\frac{Q^2}{\nu^2}) + (2\gamma_E - ln\,4 - \frac{1}{2})Q^2 \\
 &&+ 2c^2ln(\frac{Q^2}{4\nu^2}) + \text{const.} + \frac{2c^4}{3Q^2} \\
  && + \frac{4c^6}{3Q^4}  + O(\frac{1}{Q^6}), \quad C_0 = \frac{R^3}{k}.
\end{eqnarray*}
which contains a nonzero 1/Q$^2$ term proportional to c$^2$. The value of c is related to the $\rho_0$ mass by interpreting eigenvalues of (\ref{eqkum}) as mass spectrum of the associated light scalar meson. \\
For the scalar gluonium correlator case, our holographic model with h(z) = e$^{-\frac{2}{3}c^2z^2}$ is equivalent to the soft wall model, so it is natural to test the universality of this choice in the present case. It is not difficult to show that the second order z-expansion of h(z) gives an equation of motion \ref{hmbeom} identical to the motion equation of the soft wall model.

\section{Conclusion}
In view of these results, one remarks that correlators estimated from AdS/QCD models depend mainly on boundary conditions of the bulk equation of motion. Apart the boundary condition on the bulk-to-boundary propagator, one needs to select an additional physical constraint in the h-model for the quark scalar current case. An example of such constraint is to impose the normalizability of solutions of the bulk motion equation, or an alternative valuable choice, used in \cite{olan10} for different context, is to match limit c=0 solutions of the model with solutions expected in the boundary theory dual to the pure AdS bulk theory. In future work, we plan to test such assumptions in the h-model to other QCD channels and to compare it with QCD spectral sum rules results. 

\section*{Acknowledgements}
\nin
I am grateful to Prof. Stephan Narison for its guidance and to give me opportunity to communicate this work during the QCD 10, the 25-th anniversary of QCD Montpellier conference series.



\end{document}